\documentclass[aps,prl,twocolumn,showpacs, superscriptaddress, groupedaddress,nofootinbib]{revtex4}  
\usepackage{graphicx}  
\usepackage{dcolumn}   
\usepackage{bm}        
\usepackage{amssymb, amsmath}
\usepackage{slashed}   
\usepackage[bookmarks]{hyperref}

\usepackage[usenames,dvipsnames,svgnames,table]{xcolor}

\def\beq{\begin{equation}}
\def\eeq{\end{equation}}

\begin{document}
\widetext
\leftline{Saclay-t17/014}
\leftline{CERN-TH-2017-032}



\title{R-axion at colliders}
\author{Brando Bellazzini}
\affiliation{Institut de Physique Th\'eorique, Universit\'e Paris Saclay, CEA, CNRS, F-91191 Gif-sur-Yvette, France}
\affiliation{Dipartimento di Fisica e Astronomia, Universit\'a di Padova, Via Marzolo 8, I-35131 Padova, Italy}
\author{Alberto Mariotti}
\affiliation{Theoretische Natuurkunde and IIHE/ELEM, Vrije Universiteit Brussel, and International Solvay
Institutes, Pleinlaan 2, B-1050 Brussels, Belgium}
\author{Diego Redigolo}
\affiliation{Raymond and Beverly Sackler School of Physics and Astronomy, Tel-Aviv University, Tel-Aviv
69978, Israel}
\affiliation{Department of Particle Physics and Astrophysics, Weizmann Institute of Science, Rehovot 7610001,Israel}
\author{Filippo Sala}
\affiliation{LPTHE, UMR 7589 CNRS, 4 Place Jussieu, F-75252, Paris, France}
\author{Javi Serra}
\affiliation{Theory Division, CERN, CH-1211 Geneva 23, Switzerland}

\begin{abstract}
\noindent We study the effective theory of a generic class of hidden sectors where supersymmetry is broken together with an approximate R-symmetry at low energy. The light spectrum contains the gravitino and the pseudo-Nambu-Goldstone boson of the R-symmetry, the R-axion. 
We derive new model-independent constraints on the R-axion decay constant for R-axion masses ranging from GeV to TeV, which are of relevance for hadron colliders, lepton colliders and B-factories.
The current bounds allow for the exciting possibility that the first sign of SUSY will be the R-axion. 
We point out its most distinctive signals, providing a new experimental handle on the properties of the hidden sector and a solid motivation for searches of axion-like particles.


\end{abstract}

\maketitle



In this letter we argue that there are generic signs of supersymmetry (SUSY) to be looked for at colliders that have not yet been satisfactorily explored: those associated with the R-axion, the pseudo Nambu-Goldstone boson (PNGB) of a spontaneously broken R-symmetry. 

Although it is well known that supersymmetry must be broken in a ``hidden sector'', its dynamics is left unspecified in the vast majority of phenomenological studies, which instead focus on the ``visible sector'', \textit{e.g.}~the MSSM.
Here we point out that in an extensive class of models where both SUSY and R-symmetry are broken at low energy, the hidden sector leaves its footprints in observables accessible to the current experimental program. In particular, we perform a thorough phenomenological study of the R-axion at high and low energy hadron and lepton colliders.

The $\mathcal{N} = 1$ SUSY algebra contains a single $U(1)_R$ (``R-symmetry'') under which supercharges transform, $[R,Q_\alpha] = - Q_\alpha$, such that components of a given supermultiplet have R-charges $r$ differing by one unit (e.g. gauge fields carry no R-charge while gauginos have $r_\lambda~=~1$). R-symmetry plays a crucial role in models of low energy dynamical SUSY breaking.
According to the general result of Nelson and Seiberg, an R-symmetry must exist in any generic, calculable model which breaks SUSY with F-terms, and if the R-symmetry is spontaneously broken then SUSY is also broken \cite{Nelson:1993nf}. Spontaneous R-symmetry breaking often occurs also in incalculable models like \cite{Affleck:1983vc,Affleck:1984mf}. If the SUSY breaking vacuum is metastable, like in ISS constructions \cite{Intriligator:2006dd}, then an analogue of the Nelson Seiberg result holds for an approximate R-symmetry~\cite{Intriligator:2007py}.
When $U(1)_R$ is explicitly broken by a suitable deformation of the hidden sector, the R-axion gets a mass in addition to the irreducible contribution from supergravity \cite{Bagger:1994hh} but can remain naturally lighter than the other hidden sector resonances.


In light of the above observations, and contrary to what previously explored in the literature, in our phenomenological study we treat the R-axion mass as a free parameter, together with its decay constant. Our analysis shows that such a particle could well be the first sign of supersymmetry to show up at experiments. We also derive model-independent bounds on the scale of spontaneous R-symmetry breaking, opening a new observational window on the properties of the SUSY-breaking hidden sector.
The intimate connection of the R-axion with the hidden sector dynamics is reflected in its sizeable decay into the Goldstone of spontaneously broken SUSY, the Goldstino.
This decay mode provides a way to distinguish the R-axion from other axion-like particles
.

\paragraph{{\bf Setup}}

The R-axion is a PNGB realizing non-linearly the spontaneously broken approximate R-symmetry of the hidden sector. Its decay constant $f_a$ and mass $m_a$ are a priori free parameters.
If we parametrize with $m_{\ast}$ the SUSY mass gap of the hidden sector and with $g_{\ast}$ the coupling strength between hidden sector states at $m_{\ast}$, the generic size of the SUSY breaking VEV is $F\sim m_{\ast}^2/g_\ast$, an outcome of naive dimensional analysis (NDA) with a single scale and coupling \cite{Cohen:1997rt,Luty:1997fk}. The R-axion decay constant is $f_a\sim m_{\ast}/g_{\ast}$, while the R-axion mass should satisfy $m_a\ll m_{\ast}$ in order for the R-axion to be a PNGB.

As a generic consequence of spontaneous SUSY breaking, a light gravitino is also present in the low energy spectrum. In the rigid limit (i.e. $M_{Pl}\to \infty$) the transverse degrees of freedom of the gravitino decouple, leaving a massless Goldstino in the spectrum. The effective action of the Goldstino and the R-axion can be written using the non-linear superfield formalism of \cite{Komargodski:2009rz} and reads
\begin{align}
&\mathcal{L}_{\rm hid}\! =\!\int\!\!  d^4\theta( X^\dagger X+\frac{f_a^2}{2} \mathcal{R}^\dagger \mathcal{R})+\!\! \int\!\! d^2\theta (F X+w_R \mathcal{R}^2)+\text{c.c.}\notag\\
&\!\!\supset\!-F^2+i\bar{G}\bar{\sigma}^\mu \partial_\mu G+\frac{f_a^2}{2}(\partial_\mu a)^2 - \frac{w_R}{F^2}\!\left[ i G^2 e^{-2ia} \Box a+\text{c.c.}\right] \label{eq:Raxionhidden}
\end{align}
where $X$ and $\mathcal{R}=e^{i\mathcal{A}}$ carry R-charge 2 and 1 respectively and satisfy the non-linear constraints $X^2=0$ and $X(\mathcal{R}^\dagger\,\mathcal{R}-1)=0$. As a solution of the first constraint, the bottom component of $X$ is integrated out in terms of the Goldstino bilinear and its F-component gets identified with (minus) the SUSY-breaking scale $F$. Analogously, all the degrees of freedom of the chiral field $\mathcal{A}$ become functions of the Goldstino and its real bottom component $a$, which we identify as the R-axion (see the Appendix for details). 

Since the R-charge of $\mathcal{R}$ is 1, its effective action differs from the one of a SUSY axion in that a superpotential term is allowed. This is controlled by the dimension three parameter $w_R$, which is related to the VEV of the superpotential and satisfies the inequality $w_R< f_a F/2\sqrt{2}$, under the assumption of no extra light degrees of freedom other than the R-axion and the Goldstino~\cite{Dine:2009sw,Bellazzini:2016xrt}. The superpotential term induces cubic interactions between the R-axion and two Goldstini, proportional to  $m_a^2$, that lead to an invisible decay channel for the R-axion.
The corresponding decay rate of the R-axion into two Goldstini is
\beq
\Gamma(a\to GG)= \frac{1}{4\pi}\left(\frac{m_a^5 w_R^2}{f_a^2 F^4}\right)< \frac{1}{32\pi}\frac{m_a^5}{F^2}
\label{eq:atoGG}
\eeq 
and it is bounded from above as a consequence of the upper bound on $w_R$, saturated only in free theories. Our power counting gives $w_R\sim F f_a$, making the width within an $\mathcal{O}(1)$ factor of the upper limit in Eq.~\eqref{eq:atoGG}. For ordinary axions $w_R$ would instead break explicitly the associated global symmetry, resulting in a suppression of the decay width into Goldstini by extra powers of $m_a^2/m_{\ast}^2$. Hence, a sizeable invisible decay width is a distinctive feature of the R-axion compared to other axion-like particles. 

The R-axion mass is generated by sources of explicit R-symmetry breaking and
can be parametrized as
\beq
m_a^2\sim \frac{\epsilon_{\slashed{R}}F}{f_a^2} r_\epsilon^2\ll m_{\ast}^2 \quad {\rm from} \quad \mathcal{L}_{\slashed{R}}\! =\!\int\!\!d^2\theta \frac{1}{2}\epsilon_{\slashed{R}}X R^{-r_\epsilon}\label{eq:Rsymmebreaking}
\eeq 
where $r_\epsilon$ is the R-charge of the explicit-breaking spurion $\epsilon_{\slashed{R}}$, with $\epsilon_{\slashed{R}}/F\ll 1$ technically natural. 
Explicit examples of this mass hierarchy arise by 
adding suitable R-symmetry breaking deformations in calculable models of dynamical SUSY breaking like the 3-2 model~\cite{Affleck:1984xz,Nelson:1993nf,new} or in SUSY QCD at large $N$ once the hidden gauginos and squarks get soft masses~\cite{Martin:1998yr,Dine:2016sgq}.
Moreover, in SUSY-breaking models
like the one in~\cite{Intriligator:2007py}, the explicit breaking of the R-symmetry is generically bounded from above ($\epsilon_{\slashed{R}}/F \ll 1$) by requiring the SUSY-breaking vacuum to be metastable.

The R-symmetry breaking contribution \eqref{eq:Rsymmebreaking} can well be expected to dominate
over the unavoidable SUGRA contribution arising from the tuning of the cosmological constant \cite{Bagger:1994hh}\footnote{
The latter arises from a constant term in the superpotential $w_{\slashed{R}}$ generated by a sequestered sector, which in the rigid limit is completely decoupled from the sector where the R-axion lives. Including gravitational interactions at the linearized level, the two sectors gets coupled via $\mathcal{L}_{\text{SUGRA}}\supset M(FX+w_R \mathcal{R}^2+w_{\slashed{R}}+\dots)$ \cite{DiPietro:2014moa} and the resulting potential after integrating out the auxiliary fields is \begin{equation*}
V_{\text{SUGRA}}= F^2-\frac{3}{M_{Pl}^2}\left(w_R^2 + w_{\slashed{R}}^2-2w_R w_{\slashed{R}}\cos 2 a\right)
\end{equation*}
Since $w_R$ is bounded from above and $f_a\ll M_{Pl}$, the flat space time is recovered by tuning the explicit R-symmetry breaking parameter $w_{\slashed{R}}\approx M_{Pl} F/\sqrt{3}$. The resulting mass for the R-axion is 
\begin{equation*}
\label{sugra}
m_a^2 \approx \frac{8\sqrt{3}w_RF}{f_a^2M_{Pl}}=\frac{24w_Rm_{3/2}}{f_a^2}\ ,
\end{equation*}
where in the second equality we used $m_{3/2}=F/\sqrt{3}M_{Pl}$ for the gravitino mass.
},
which gives rise to $m_a^2\sim (1\text{ MeV})^2\times \frac{m_{\ast}}{10\text{ TeV}}\times \frac{m_{3/2}}{0.01\text{ eV}}$.

We are now ready to study the couplings of the R-axion with the visible sector fields, which we take to be the MSSM (with matter-/R-parity). The superpartners get SUSY-breaking masses $m_{\text{soft}}$ from their interactions with the hidden sector, which are controlled by a perturbative coupling $g$. This coupling is a proxy for the SM gauge coupling constants in gauge mediation models \cite{Giudice:1998bp,Meade:2008wd} or for Yukawa-type interactions in extended gauge mediation, see \cite{Evans:2013kxa} for a review. The scaling of $m_{\text{soft}}$ strongly depends on the type of mediation mechanism. We can estimate it as 
\begin{equation}
m_{\text{soft}}\sim\left(\frac{g}{g_\ast}\right)^n \times g \frac{F}{m_\ast}=\left(\frac{g}{g_\ast}\right)^{n+1}\times m_{\ast}
 \label{eq:msoft_scaling}
\end{equation}
In this letter we assume that gauginos get a mass via their coupling to the hidden sector global current, so that $m_{\text{soft}} \sim (g/g_\ast)^2 m_{\ast}$. Notice that if $g_\ast=4\pi/\sqrt{N_{\text{mess}}}$ we recover the ordinary gauge mediation scaling where $N_{\text{mess}}$ is the number of messengers.
Other scaling, \textit{e.g.}~the one of \cite{Gherghetta:2011na} for Dirac gauginos ($n=0$) will be discussed elsewhere.
Besides,
whatever the scaling in eq.~(\ref{eq:msoft_scaling}), there is always a large portion of parameter space where the R-axion is lighter than the superpartners, which correspond to $r_\epsilon\sqrt{g_*\epsilon_{\slashed{R}}}/m_\ast\lesssim (g/g_\ast)^{n+1}$. 
It would also be interesting to depart from the NDA expectation for the scales $F$ and $f_a$ and explore models where a large separation between the two is realized. 

We consider in the following a small SUSY breaking scale $\sqrt{F}$ in the range from 1 to a few 10's of TeV. This regime is welcome for fine-tuning and Higgs mass considerations. The resulting gravitino mass lies in the window $10^{-4}\text{ eV}\lesssim m_{3/2}\lesssim 5\text{ eV}$, where the upper limit comes from cosmological and astrophysical bounds on gravitino abundance \cite{Pierpaoli:1997im,Viel:2005qj,Osato:2016ixc}, while the lower limit comes from collider bounds on gravitino pair production in association with a photon or a jet at LEP~\cite{Brignole:1997sk} and at the LHC~\cite{Brignole:1998me,Maltoni:2015twa}.

Since the visible sector feels the SUSY-breaking only through $g/g_{\ast}$ effects, we can treat the MSSM superfields linearly and ``dress'' the R-charged operators with appropriate powers of the R-axion. We also neglect subleading effects in the explicit R-breaking, suppressed by powers of $\sim m_a/m_{\ast}$. 
The interactions of the R-axion with the MSSM gauge sector are then
\begin{align}
&\mathcal{L}_{\text{gauge}}=\int\!\! d^2\theta \left[-i g_i^2\,\frac{c_i^{\text{hid}}}{16 \pi^2} \mathcal{A}\,-\,\frac{m_{\lambda_i}}{2 F} X \mathcal{R}^{-2}\,\right]\mathcal{W}_i^2+ \rm c.c.\notag\\
&\supset\!  \frac{ g_i^2c_i^{\text{hid}}}{16 \pi^2}\! \frac{a}{f_a} F^i \tilde{F}^{i} -\frac{m_{\lambda_i}}{2} \lambda_i\lambda_i\, \left[e^{-2i a/f_a}\! +\! \frac{g_i^2 c_i^{\text{hid}}}{4 \pi^2} i\frac{a}{f_a}\right]+ \rm c.c.\,
\end{align}
where $\mathcal{W}$ is the field strength superfield carrying R-charge 1 and $i$ labels the SM gauge group, where we defined $g_1 = \sqrt{5/3}\,g_Y$ and $\tilde{F}^{i,\mu\nu}=1/2\epsilon^{\mu\nu\rho\sigma} F_{\rho\sigma}^i$. The Majorana gaugino masses are of order $m_{\lambda_i}\approx m_{\text{soft}}$ by assumption. The coefficients $c_i^{\text{hid}}$ encode the hidden sector contributions to the mixed anomalies of the $U(1)_R$ with the SM gauge groups.
For example, we get $c_i^{\text{hid}}=-N_{\text{mess}}$ for $i=1,2,3$, for $N_{\rm mess}$ messengers chiral under $U(1)_R$ and in the $5+\bar{5}$ of $SU(5)$ with zero R-charge (in our NDA $N_{\text{mess}} \sim (4\pi/g_*)^2$). The contributions to the anomalies from the MSSM fields will be encoded in the full loop functions. 

The interactions in the Higgs sector can be written as
\begin{align}
&\mathcal{L}_{\text{higgs}}=\!\int\!\! d^4\theta\! \left[\frac{\mu}{F}X^\dagger\mathcal{R}^{2-r_H}
-\frac{B_\mu}{F^2}\,\vert X\vert^2 \mathcal{R}^{-r_H}\right]\! H_u H_d +{\rm c.c.}  \notag\\
&\supset\,\mu\,\tilde{h}_u \tilde{h}_d e^{i (2-r_H ) a/f_a}-B_\mu h_u h_d\, e^{- i r_H a/f_a}+{\rm c.c.}
\end{align}
where $\tilde{h}_{u,d}$ are the Higgsino Weyl spinors and $h_{u,d}$ the complex Higgs scalar doublets.
We have assumed the $\mu$-term to be generated by the hidden dynamics\footnote{A $\mu H_u H_d$ term in the superpotential would break $U(1)_R$ explicitly for $r_H \neq 2$, yielding $m_a \lesssim \mu\,m_{\rm soft}/4 \pi f_a \approx O(100)\text{ GeV} \cdot (\mu\,m_{\rm soft}/{\rm TeV}^2) \cdot ({\rm TeV}/f_a)$.
}, so that the total R-charge of the Higgses $r_H = r_{H_u} + r_{H_d}$ depends on the charge assignments in the sector responsible for generating $\mu$ and $B_\mu$. The charge assignment of the visible sector fields is modified by higher dimensional operators in the Kahler like $\vert H_{u,d}\vert^2 \vert \mathcal{R}\vert^2$, etc., which lead to $g/g_{\ast}$ suppressed effects that will be neglected in what follows. Notice also that the NDA size of $\mu\sim g^2  m_{\ast}/g_{\ast}^2$ and $B_\mu\sim g^2 m_{\ast}^2/g_{\ast}^2 $ reflects the well known $\mu-B_\mu$ problem in low energy SUSY-breaking scenarios. 
 
The coupling to the MSSM Higgses proportional to $B_\mu$ induces, after electroweak symmetry breaking, a small mixing between $a$ and the MSSM Higgs boson $A$ \footnote{Eq.~\eqref{eq:delta} accounts for both mass and kinetic mixing between $A$ and $a$. In fact, in the limit $m_a\to0$, $\delta$ parametrizes the misidentification of $a$ as the R-axion after EWSB (given that for $r_H \neq 0$ the Higgs is R-charged, $f_a^2\to f_a^2+r_H v^2s^2_{2\beta}$).}
\beq
\delta = r_H \frac{v}{f_a} \frac{s_{2 \beta}}{2} \frac{1}{1 - m_a^2/m_A^2}\simeq r_H \frac{v}{f_a}\! \frac{s_{2 \beta}}{2}\label{eq:delta}
\eeq
If we assume the Yukawa interactions in the superpotential to be allowed in the limit of exact $U(1)_R$ ($r_{H_u} + r_Q + r_U = 2$, etc.), the mixing $\delta$ is the only source of couplings between $a$ and the SM fermions and we get
%
\begin{equation}
\mathcal{L}_{\text{f}}= i r_H\frac{a}{f_a}\! \left[c_\beta^2\,m_u\bar{u}  \gamma_5 u +\! s_\beta^2\,m_d\bar{d}  \gamma_5 d + s_\beta^2\,m_\ell\bar{\ell}  \gamma_5\ell\right]
\end{equation}
The same mixing induces
\begin{equation}
\mathcal{L}_{ahh}=\frac{\delta^2}{v} h (\partial_\mu a)^2 
\label{eq:atohh}
\end{equation}
where $h$ is the SM-like Higgs, as well as extra interactions with the MSSM Higgses whose phenomenological consequences we leave for future work \cite{new}.
Finally, the $a$ couplings to sfermions also arise from its mixing with $A$ and are proportional to the A-terms. Since we assume all the sfermions to be heavy and the A-terms to be small, these couplings do not play any role in the R-axion phenomenology discussed here.


\begin{figure*}[t!]
\includegraphics[scale=0.65]{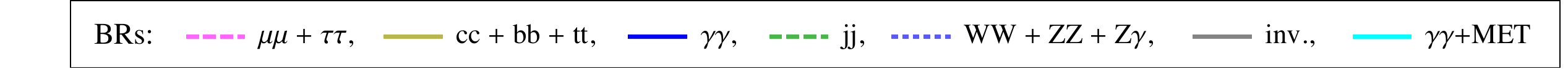}\\ \vspace{.15 cm}
\includegraphics[scale=0.65]{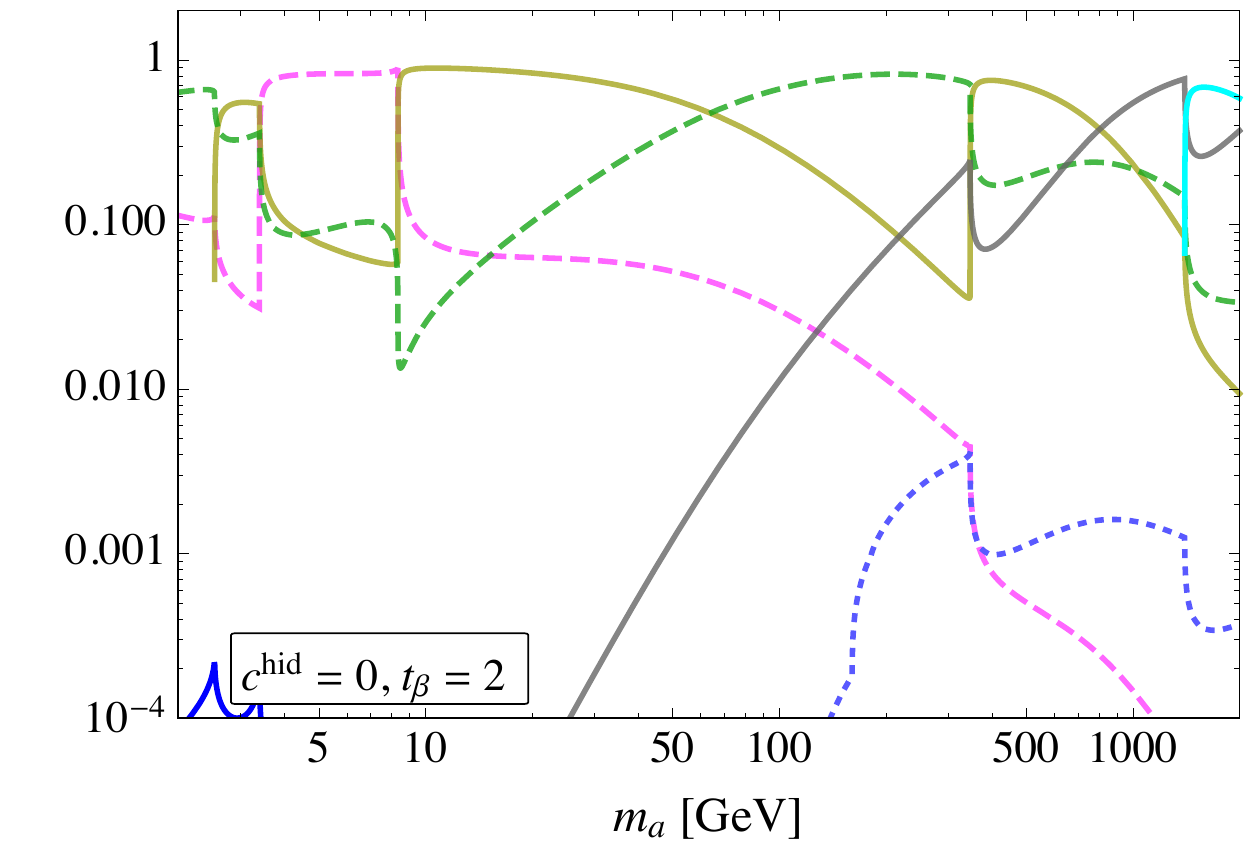} \qquad
\includegraphics[scale=0.65]{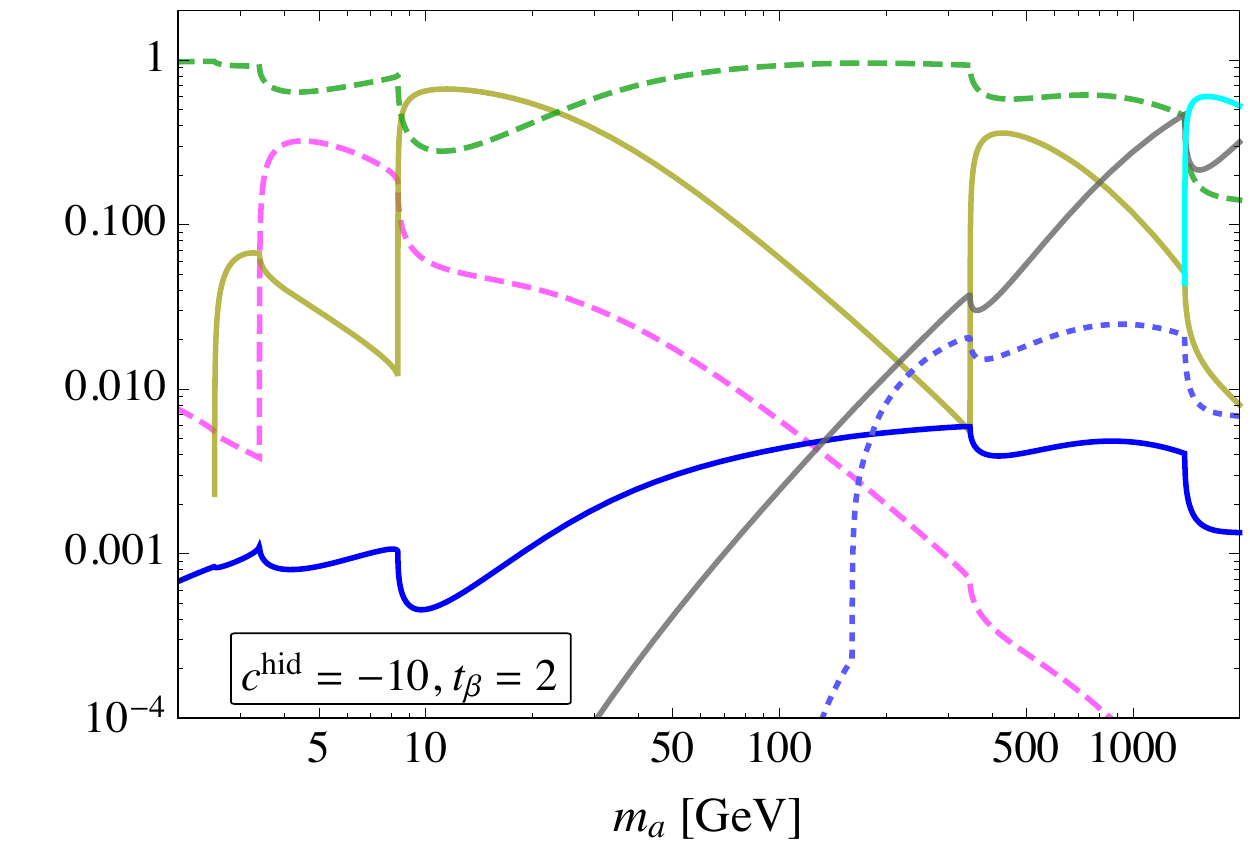}
\caption{
\label{BRRaxion}
Branching ratios as a function of the R-axion mass for two representative values of the hidden sector anomaly coefficients $c^{\rm hid}$. The relative size of the BRs does not depend on $f_a$. For the invisible decay into Goldstini we assume the inequality in Eq.~\eqref{eq:atoGG} to be saturated and we take $f_a/\sqrt{F}=0.9$.}
\end{figure*}

Finally, 
we note that all the couplings derived via the spurion analysis can be thought as derivative couplings of the R-axion with the R-current $j_R$, up to anomalous and explicit breaking terms.

\paragraph{ {\bf Phenomenology}}
We now discuss the phenomenological implications of the R-axion. We focus on R-axion masses in the range between $2\text{ GeV}$ and $2\text{ TeV}$, and we refer to \cite{Goh:2008xz} for an LHC study for masses of $O(100)$ MeV.\footnote{A much lighter R-axion would be similar to traditional axion-like particles (see \cite{Kim:2008hd} for a review):  for $r_H=2$ the R-axion couplings to fermions are the ones of the DFSZ axion \cite{Dine:1981rt,Zhitnitsky:1980tq}, while for $r_H=0$ the couplings to fermions are zero at the tree level and the phenomenology is dominated by the couplings to gluons and photons like in the KSVZ model \cite{Kim:1979if,Shifman:1979if}. The phenomenological study of the low mass window is left for future work.}
For definiteness we fix $r_H=2$, which allows for an R-symmetric $\mu$ term and a $B_\mu$ term from spontaneous $U(1)_R$ breaking.  We will comment on the phenomenological differences of the $r_H=0$ case, where the role of $\mu$ and $B_\mu$ is reversed. The Majorana gaugino masses cannot be arbitrarily larger than the scale of spontaneous R-breaking so we take $f_a\gtrsim 0.3 \text{ TeV}$ and fix for illustrative purposes the gaugino masses to the GUT universal values $m_{\lambda_{1,2,3}}=0.7, 1.4, 3.6$ TeV (different values do not change the R-axion phenomenology as long as $m_{\lambda_{i}}>m_a/2$). For $f_a\lesssim 1\text{ TeV}$, obtaining such heavy gauginos present model building challenges which are beyond the scope of this paper.   

We now discuss the different production modes of the R-axion at the LHC, at LEP and at B-factories. For the purposes of this paper we ignore R-axion production from SUSY decay chains. As for any axion-like particle, the single production modes scale with $1/f_a^2$ and double production ones with $1/f_a^4$:
\begin{itemize}
\item[$\circ$] At the LHC, the resonant $a$ (+ SM) production is dominated by gluon fusion. To determine $\sigma_{gg \to a}$ we use the leading order prediction at 13 TeV (including the hidden sector anomaly and the full  loop functions for gluino, top and bottom) multiplied by a constant $K$ factor of 2.4~\cite{Ahmed:2015qda}. For $f_a = 1$ TeV, $m_a = 100\text{ GeV}$ and $N_{\text{mess}}=10$ we get $\sigma_{gg \to a}\approx 10^2\text{ pb}$ while for $N_{\text{mess}}=0$ $\sigma_{gg \to a}\approx 20\text{ pb}$.

\item[$\circ$] Also at the LHC, we have double $a$ production from Higgs decays which is driven by the $h (\partial_\mu a)^2$ coupling in \eqref{eq:atohh}. The BR$(h\to aa)$ goes up to $10\%$ for $t_\beta=2$ and $f_a = 1$~TeV.\footnote{Even if potentially important we do not discuss here other pair production mechanisms.}  

\item[$\circ$] At LEP the R-axion can be produced via its coupling to the $Z$ boson. At LEP~I we  consider on-shell $Z$ production which then decays to $a\gamma$ with BR$(Z\to a\gamma)\approx3\times 10^{-7}$ for $f_a = 1\text{ TeV}$ and $N_{\text{mess}}=10$.  At LEP~II we consider the associated production of $\gamma a$ or $Z a$ from an off-shell $Z$ and $\gamma$. These cross sections are around $10^{-4}\text{ pb}$ for $N_{\text{mess}}=10$.

\item[$\circ$] Flavor experiments can constrain the R-axion parameter space for $m_a\lesssim10\text{ GeV}$. In particular, for $m_a\gtrsim 2\text{ GeV}$ we consider R-axion emission in $B \to K a, K^*a$ transitions and $\Upsilon \to \gamma a$ decays.
The $\text{BR}(B\to K^{(*)}a)$ are computed from the general result of \cite{Hall:1981bc}, accounting for the mixing of the R-axion with the CP-odd Higgs \eqref{eq:delta} \cite{Freytsis:2009ct}, and choosing for reference $m_{H^\pm} = 1$ TeV (we take the form-factor relevant for $K^*$ from~\cite{Ball:2004rg}).
This yields, for both $K$ and $K^*$ final states, $\text{BR} \approx 3-5\times10^{-4}$ for $f_a = 1\text{ TeV}$ and $t_\beta=2$.
The $\text{BR}(\Upsilon\to \gamma a)/\text{BR}(\Upsilon\to ll)$ is computed using the standard Wilczeck formula \cite{Wilczek:1977zn}. This simple estimate neglects the mixing of the R-axion with $\eta_b$ mesons and it is reliable for $m_a\lesssim 9\text{ GeV}$. For $f_a=1\text{ TeV}$ the $\text{BR}(\Upsilon\to \gamma a)$ is around $3-5\times 10^{-5}$. 
\end{itemize}

 \begin{figure*}[t]
\includegraphics[scale=0.65]{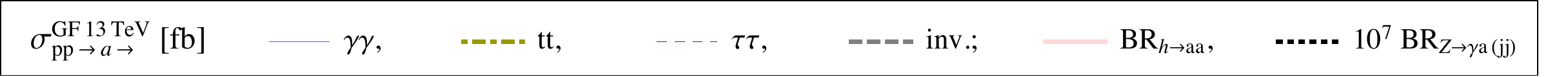}\\ \vspace{.15 cm}\hspace{0.02 cm}
\includegraphics[scale=0.55]{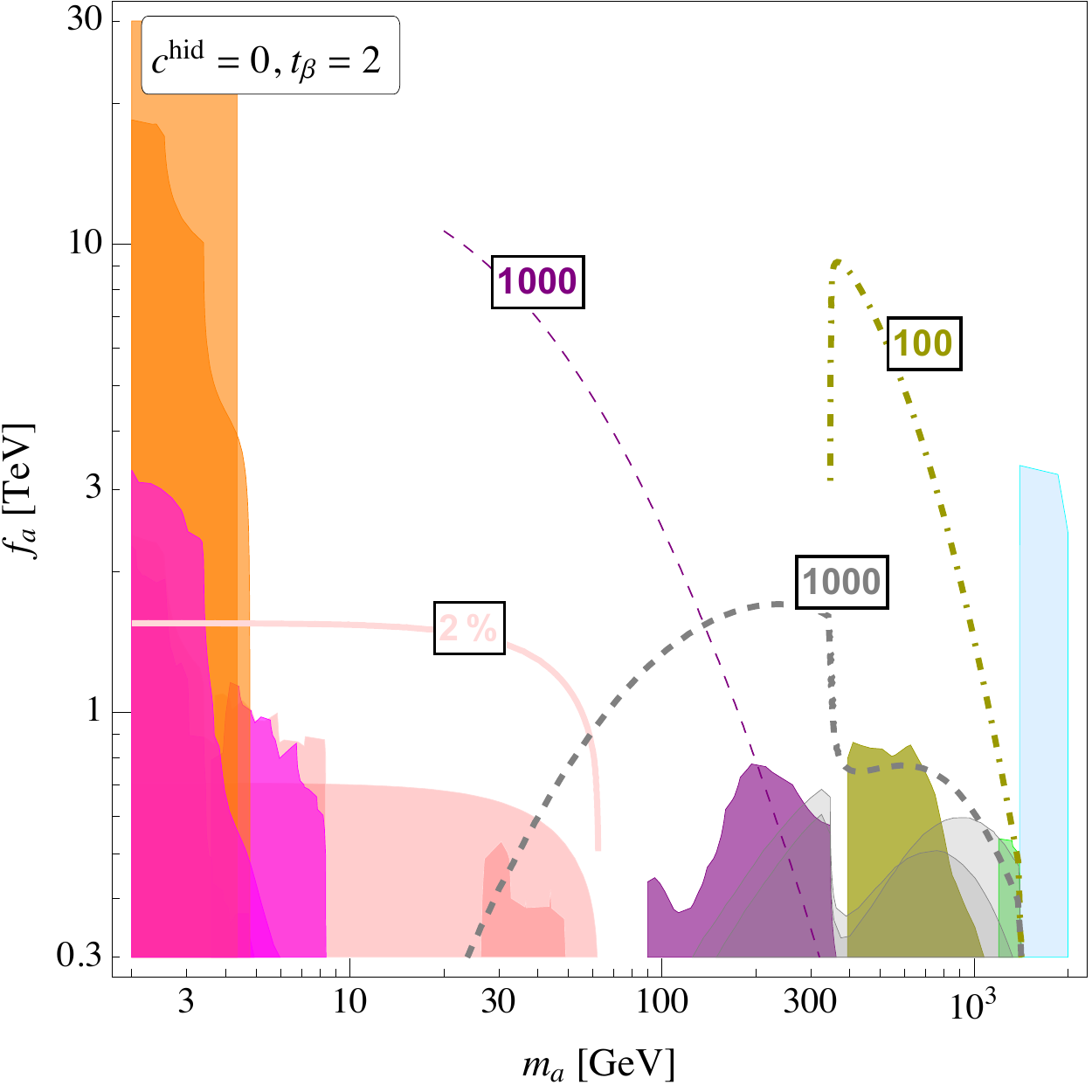} \qquad \includegraphics[scale=0.55]{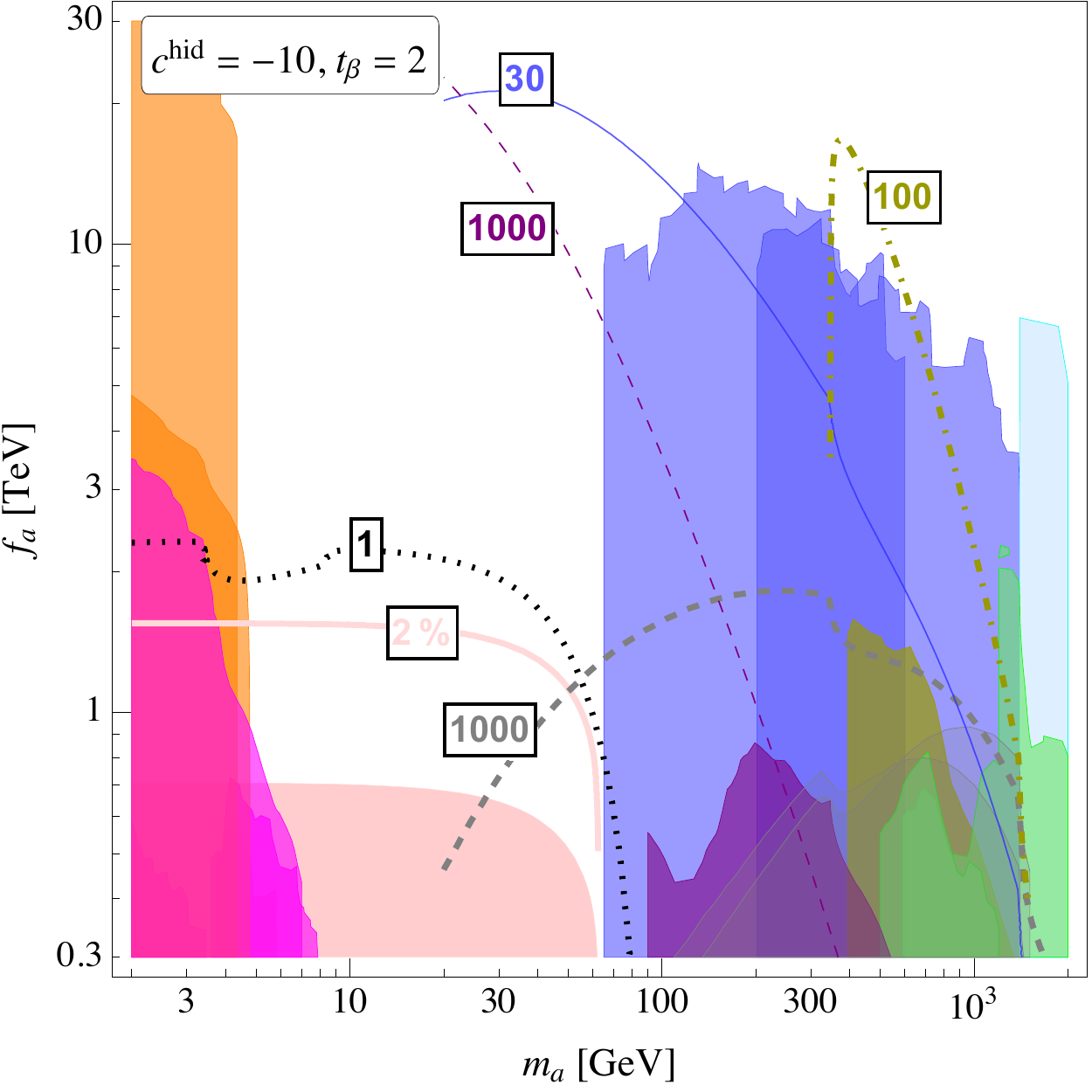}
\quad\includegraphics[scale=0.6]{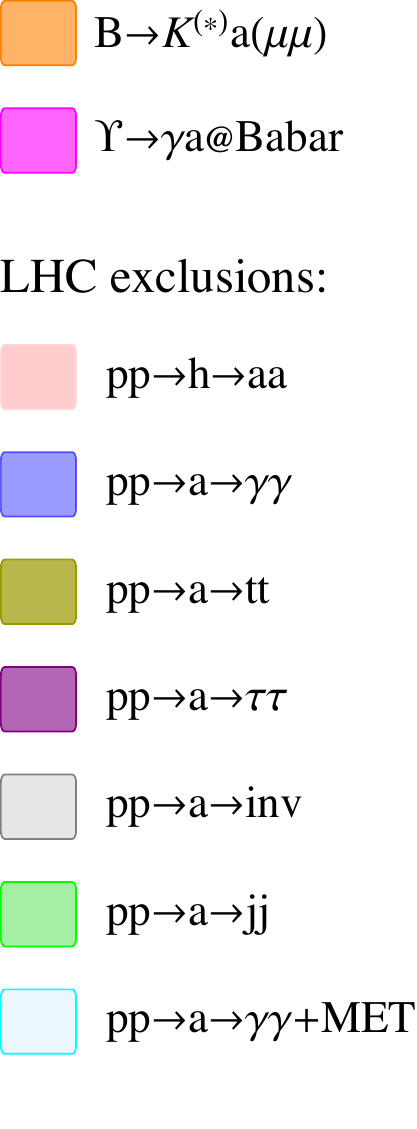}
\caption{\label{fig:ma_fa} Shaded: LHC8 and LHC13~\cite{Aad:2015zva,Aaboud:2016tnv,
Aad:2015hea,
Aad:2014ioa,ATLAS-CONF-2016-059,Khachatryan:2016yec,
Khachatryan:2016ecr,Sirunyan:2016iap,ATLAS-CONF-2016-069,
Aad:2015fna,Chatrchyan:2013lca,
ATLAS-CONF-2016-085,CMS:2016rjp,
Khachatryan:2016vau,
Khachatryan:2017mnf,Aad:2015oqa
}, LHCb~\cite{Aaij:2015tna,Aaij:2012vr}, Babar~\cite{Lees:2012te,Lees:2012iw,Lees:2011wb} and Belle~\cite{Hyun:2010an} exclusions. Contours: signal strengths at the LHC13 and Higgs and $Z$ boson branching ratios.
The tiny area in the lower right corner where $m_a\approx 4\pi f_a$ lies beyond the regime of validity of our effective description. 
}
\end{figure*}

In Fig.~\ref{BRRaxion} we compare the branching ratios of the R-axion in two extreme cases for the GUT anomaly coefficients $c^{h}_i=-N_{\text{mess}}$. These are modified by the loop functions of the MSSM fields, which in the limit of a light R-axion are encoded in a shift of the anomaly coefficients, \textit{e.g.}~$\Delta c_3=3 -\frac{N_f r_H}{2}$ where $N_f$ stands for the number of SM families heavier than $m_a$.
Notice that the contribution from gaugino loops partially cancels the negative one from $N_{\text{mess}}$. For $r_H=2$ and within our range of R-axion masses the decay widths into gluons, di-bosons and di-photons for $N_{\text{mess}}=0$ are suppressed with respect to the case with $N_{\text{mess}}=10$. In particular the reduced color anomaly explains the smaller cross-section from gluon fusion for $N_{\text{mess}}=0$. 

In Fig.~\ref{fig:ma_fa} we summarize the present constraints on the R-axion in the $m_a-f_a$ plane as well as the most promising processes to search for it at future experiments. For $m_a\gtrsim m_h/2$, the most important bounds come from resonant $a$ production at the LHC.
The most distinctive feature of the R-axion is the large invisible signal strength, which is enhanced at large $m_a$ because of the enhanced BR into Goldstini (see Fig.~\ref{BRRaxion} and Eq.~\eqref{eq:atoGG}). As a consequence current monojet searches from 8 and 13 TeV data \cite{Aad:2015zva,Aaboud:2016tnv} constrain the R-axion parameter space.  To draw the monojet exclusions we have determined the ratio of the $a$ and $a$ + jet(s) production cross sections via a MadGraph~\cite{Alwall:2011uj,Alwall:2014hca} simulation, for the different missing energy cuts for which the bounds are given in \cite{Aad:2015zva,Aaboud:2016tnv} (we believe this approximation to be sufficient for our purposes).

At $m_a>1.4\text{ TeV}$, we show how the decay into bino pairs opens up. Given the light gravitino, the bino promptly decays to $\gamma+G$, resulting in a $\gamma\gamma+\text{MET}$ final state which is constrained by inclusive $\gamma\gamma+\text{MET}$ searches \cite{Aad:2015hea}. We translate such searches in a bound $\sigma_{gg\to a\to \lambda_1\lambda_1}^{8 \rm TeV} < 0.3$~fb, which results in $f_a\gtrsim 6\text{ TeV}$ for $N_{\text{mess}}=10$ and $f_a\gtrsim 3\text{ TeV}$ for $N_{\text{mess}}=0$, because of the reduced production cross section.

For large anomalies the dominant decay mode is in di-jets for the full R-axion mass range (we included a constant $K$-factor of 1.5~\cite{Djouadi:2005gj}).
The branching ratio into di-photons is always around $1/8\,\left(2\alpha_{\text{em}}/\alpha_{\text{s}}\right)^2\sim 0.1\%$ and the di-photon resonant searches  \cite{Aad:2014ioa,ATLAS-CONF-2016-059,Khachatryan:2016yec} at 8 and 13 TeV dominate the collider phenomenology for $m_a>60\text{ GeV}$~\footnote{We checked that the decays into di-bosons and $Z\gamma$ can be neglected for our choice of the anomalies.}. The net result is a lower bound on $f_a$ around 10 TeV over the full mass range  for $m_a\gtrsim m_h/2$.
Searches for a resonance decaying into di-jet \cite{Khachatryan:2016ecr,Sirunyan:2016iap,ATLAS-CONF-2016-069}, into $t\bar{t}$  at 8 TeV \cite{Aad:2015fna,Chatrchyan:2013lca} and into di-tau at 13 TeV \cite{ATLAS-CONF-2016-085,CMS:2016rjp} give complementary bounds for $m_a>500\text{ GeV}$, $m_a>2 m_t$, $100\text{ GeV}\lesssim m_a<2 m_t$ respectively.
Notice that the $R$-axion is typically narrow, \textit{e.g.} $\Gamma_a/m_a \approx 10^{-3}$ for $m_a = 1$~TeV, $f_a = 3$~TeV and $N_{\rm mess} = -10$.

For small anomalies the LHC constraints are sensibly weaken by the reduced production cross section and by the suppressed branching ratio in di-photons. A lower bound on $f_a$ can still be derived from a combination of $t\bar{t}$, di-tau and monojet searches. In particular the monojet searches give the dominant constraints for $m_a\gtrsim700\text{ GeV}$ and are competitive with di-taus searches for $m_a<2 m_t$.

For $m_a\lesssim 2 m_h$, the major constraint comes from the upper bound on $\text{BR}(h\to\text{untagged})<32\%$ \cite{Khachatryan:2016vau}, which can be translated in a lower bound on $f_a\gtrsim 700\text{ GeV}$. This bound depends only on the mixing in Eq. \eqref{eq:delta} and applies to  both cases with $N_{\text{mess}}=10$ and $N_{\text{mess}}=0$. We also include constraints arising from exclusive Higgs decays (\textit{e.g.} $h\to aa\to 4\mu$ or $h\to aa\to 2\tau2\mu$) \cite{Khachatryan:2017mnf,Aad:2015oqa}. 
Finally, LEP constraints~\cite{Abreu:1994du,Acciarri:1994gb,Rupak:1995kg,Adriani:1992zm,Acciarri:1994ye,Anashkin:1999da} are not relevant for the $f_a$ considered in this study.

For $m_a\lesssim 9\text{ GeV}$, stringent constraints on $f_a$ come from Babar searches on $\Upsilon\to a\gamma$ with $a$ decaying into tau or muon pairs \cite{Lees:2012te,Lees:2012iw} or hadrons \cite{Lees:2011wb}. These give a bound $f_a\gtrsim1\text{ TeV}$ which goes up to 3 TeV for both $N_{\text{mess}}=10$ and $N_{\text{mess}}=0$.
For $m_a\lesssim 4\text{ GeV}$, the stronger bound on $f_a$ is given by the LHCb latest result $\text{BR}(B\to K^{\ast}\mu\mu)\lesssim2\times10^{-9}$ \cite{Aaij:2015tna}, but we also considered bounds from Belle and older LHCb data \cite{Hyun:2010an,Aaij:2012vr}. This results in $f_a\gtrsim30\text{ TeV}$ for $N_{\text{mess}}=10$ which goes up to $\approx 100$ TeV for $N_{\text{mess}}=0$, where the $\text{BR}(a\to\mu\mu)$ is enhanced.

Notice that different values of $t_\beta$ modify the value of the mixing $\delta$ in Eq.~\eqref{eq:delta}. In particular $\delta$ becomes smaller at large $t_\beta$ reducing the bounds from Higgs branching ratios measurements and $B\to K$ transitions. The couplings to quarks and leptons are also $t_\beta$ dependent, most importantly the $t \bar{t}$ signal strength is reduced at large $t_\beta$. For $r_H=0$ (and within our previous assumptions) the R-axion does not couple to SM fermions at the linear level.
This makes it generically very difficult to be constrained for $m_a<m_h/2$.
For larger values of $m_a$ and irrespectively of the values of $t_\beta$ and $r_H$, diphoton constraints give $f_a\gtrsim 10~\text{TeV}$ for large anomalies, while for small anomalies a milder bound on $f_a$ is anyway given by monojet and di-jet searches.

We now discuss the relevant signatures for future discovery of the R-axion. The most distinctive one is the decay into two Goldstini which gives a large invisible signal strength (dashed grey lines in Fig~\ref{fig:ma_fa}). This will be probed by monojet searches at the LHC (multijet+MET searches could also be relevant~\cite{Buchmueller:2015eea}) and constitutes a very good motivation for the high-luminosity LHC program.

Other promising signatures for the future experimental programs are shown in Fig.~\ref{fig:ma_fa}. For large anomalies di-photon will be the most promising final state at the LHC, while for small anomalies di-taus and $t\bar{t}$ will be more important. For $m_a<m_h/2$ we show how an improvement of the Higgs coupling measurements down to $1-2\%$ (which is within the reach of ILC \cite{Dawson:2013bba}) would probe $f_a$ up to 1.5 TeV. Even bigger values of $f_a$ are within the reach of machines like CLIC, CEPC and FCC-ee which plan to probe Higgs coupling with a precision of roughly $10^{-3}$~\cite{Dawson:2013bba}.
For large anomalies, an important probe of a light R-axion would be $Z\to\gamma a$ measurements at future lepton colliders.
A naive rescaling of the LEP~I analysis~\cite{Adriani:1992zm}, for example, indicates that $Z\to\gamma a(jj)$ BRs in the ballpark of $10^{-7}$ could be probed at the FCC-ee, if $O(10^{12})$ $Z$'s will be produced.
We notice that the mass window $10~{\rm GeV} \lesssim m_a \lesssim 65$ GeV is less constrained by the searches we considered.
This could be improved by extending the coverage of resonance searches, in particular $\gamma\gamma$, to lower invariant masses.

To distinguish the R-axion from other scalar resonances, a $\text{jet}+\text{MET}$ signal would certainly help in combination with a pattern along the lines discussed above. Of course, to reinforce the R-axion interpretation of a possible signal, one would eventually need to find evidence for superpartners.

\paragraph{ {\bf Conclusions}}
The possibility that the R-axion could be the first sign of SUSY at colliders is well motivated from theoretical as well as phenomenological considerations. 

In this letter we have investigated the low energy dynamics of SUSY breaking sectors with a light R-axion (and gravitino) coupled to the MSSM.
Our results are summarised in Fig. 2, where we show how current and future colliders probe the space of R-axion masses and decay constants. 
We have also identified some promising signatures to cover the currently unconstrained part of the parameter space.

The R-axion constitutes a very interesting prototype of axion-like particles, 
with couplings that follow from well defined selection rules
of the theory, and whose mass can be safely considered a free parameter.

The rich phenomenology of the R-axion certainly deserves further investigation. The R-axion can give rise to non-standard heavy Higgs decays or SUSY decay chains, it can be a further motivation for high intensity experiments (in its light mass window), and could impact cosmological and astrophysical processes. 

Finally, we wish to point out that other appealing features of SUSY, such as unification and dark matter, might find an interesting interplay with a light R-axion, opening new model building avenues.
We leave the exploration of this exciting physics for the future.

\medskip

\subsection{Acknowledgements}

We thank Lorenzo Di Pietro, Zohar Komargodski, David Shih, Riccardo Torre and Lorenzo Ubaldi for useful discussions. The authors thank CERN and the LPTHE for kind hospitality during the completion of this work.
F.S. is grateful to the Institut d'Astrophysique de Paris ({\sc Iap}) for hospitality.
\smallskip

{\footnotesize
\noindent Funding and research infrastructure acknowledgements: 
\begin{itemize}
\item[$\ast$] B.B. is supported in part by the MIUR-FIRB grant RBFR12H1MW ``A New Strong Force, the origin of masses and the LHC'';
\item[$\ast$] A.M. is supported by the Strategic Research Program High Energy Physics and the Research Council of the Vrije Universiteit Brussel;
\item[$\ast$] F.S is supported by the European Research Council ({\sc Erc}) under the EU Seventh Framework Programme (FP7/2007-2013)/{\sc Erc} Starting Grant (agreement n.\ 278234 --- `{\sc NewDark}' project).
\end{itemize}
}

\subsection*{Appendix}
For convenience of the reader, we report here the explicit expressions of the constrained superfields $X$ and $\mathcal{R}$, that satisfy $X^2=0$ and $X(\mathcal{R}^\dagger\,\mathcal{R}-1)=0$~\cite{Komargodski:2009rz,Dine:2009sw}
 \beq
X=\dfrac{G^2}{2F_X}+\sqrt{2}\theta G+ \theta^2 F_X
\eeq
 \beq
\mathcal{R} 
= e^{i {\mathcal A}}, \quad \mathcal{A} = \tilde{a} +\sqrt{2}\theta \psi_\mathcal{R} + \theta^2 F_\mathcal{R}
\eeq
 \beq
\tilde{a} = a - \dfrac{i}{2}\,\left(\dfrac{G}{F_X}\sigma^\mu \dfrac{\bar{G}}{\bar{F}_X}\right) \partial_\mu a+ \cdots \label{eq:atilde}
\eeq
 \beq
\psi_\mathcal{R} =- i\, \sigma^\mu \dfrac{\bar{G}}{\bar{F}_X} \partial_\mu \tilde{a}
\eeq
 \beq
F_\mathcal{R} = -\left(\partial_\nu \dfrac{\bar{G}}{\bar{F}_X}\right)\bar{\sigma}^\mu\,\sigma^\nu\,\dfrac{\bar{G}}{\bar{F}_X}\,\partial_\mu \tilde{a}
- \dfrac{1}{2} \dfrac{\bar{G}^2}{\bar{F}_X^2}\,\Box \,\tilde{a}\,
\eeq
From Eq.~(\ref{eq:Raxionhidden}) one has $F_X=-F + \cdots$, where the dots here and in Eq.~(\ref{eq:atilde}) stand for terms with more derivatives and fermions.
Notice that our convention for the metric tensor is $\eta^{\mu \nu} = {\rm diag}(+1, -1, -1, -1)$, and the one for $\sigma^\mu$, $\bar{\sigma}^\nu$ is defined by $\sigma^\mu = (\sigma^0, \sigma^i)$ and $\bar{\sigma}^\nu = (\sigma^0, -\sigma^i)$, where $\sigma^{i=1,2,3}$ are the Pauli matrices and $\sigma^0 = {\rm diag}(+1, +1)$ (these conventions imply $\{\sigma^\mu,\bar{\sigma}^\nu\} = 2 \eta^{\mu\nu}$).

\medskip

\bibliography{Raxion}
\end{document}